\documentclass{aastex}
\usepackage{spr-astr-addons}
\usepackage{mathtools}

\RequirePackage{color}

\begin{document}
\title{Gravitational lenses in the dark Universe}
\author{R. C. Freitas} \author{S. V. B. Gon\c{c}alves} \and \author{A. M. Oliveira}
\affil{Universidade Federal do Esp\' \i rito Santo, Centro de Ci\^encias Exatas, Departamento de F\' \i sica}
\affil{Av. Fernando Ferrari, 514 – Campus de Goiabeiras, CEP 29075-910, Vit\'oria, Esp\' \i rito Santo, Brazil
\\email: rc$\_$freitas@terra.com.br\\email: sergio.vitorino@pq.cnpq.br\\email: adriano.ufes@gmail.com}
\par
\begin{abstract} We discuss how different cosmological models of the Universe affect the probability that a background source has multiple images related by an angular distance, \textit{i.e.}, the optical depth of gravitational lensing. We examine some cosmological models for different values of the density parameter $\Omega_i$: i) the cold dark matter model, ii) the $\Lambda$CDM model, iii) the Bose-Einstein condensate dark matter model, iv) the Chaplygin gas model, v) the viscous fluid cosmological model and vi) the holographic dark energy model by using the singular isothermal sphere  (SIS) model for the halos of dark matter. We note that the dependence of the energy-matter content of the universe profoundly modifies the frequency of multiple quasar images.
\end{abstract}


%\section*{}
%\label{sec:intro}

%%%%%%%%%%%%%%%%%%%%%%%%%%%%%%%%
\section{Introduction}
%%%%%%%%%%%%%%%%%%%%%%%%%%%%%%%%

During the last years strong evidences for an accelerated expansion of the Universe has been found through several independent cosmological tests \citep{rie,per,wmap}. On the other hand, dynamical estimations of the amount of matter in the Universe seem to indicate the picture provided by the standard cold dark matter (CDM) scenario \citep{ost}. The combination of these evidences leads to the so called dark sector of the Universe, whose essential nature is still unknown. Actually there is a great number of cosmological models that  try to account the dark sector of the Universe. The most known are: $\Lambda$CDM \citep{arm}, quintessence cosmological model \citep{cal}, Chaplygin gas model \citep{kam}, viscous fluid cosmological model \citep{vis01,vis02,kre}, holographic dark energy model \citep{cam}, etc. Each one solves some problems but creates other questions. A possible way to improve these models and to shed light on these questions is to test them against the available observational cosmological data. The confrontation between theoretical models and observational data enable us to constraint the cosmological parameters, which is the greatest goal of the modern cosmology. There are some tools that can be used for in this aim: the distance measurements of type Ia supernovae \citep{mg}; the power spectrum fluctuations in the cosmic microwave background radiation \citep{ber, ben}, nucleosynthesis constraints \citep{tur} and so on.
\par
The gravitational lens can be other important tool for determining the cosmological parameters of our Universe. Einstein's General Theory of Relativity predicts that a massive object curves space-time in its vicinity. As a consequence of this curvature, the light emitted from a background source is deflected and its image is distorted when the light passes near massive objects, such as galaxies and galaxy clusters. The lens effect can distort and magnify the image of the source. Thus, the gravitational lensing effect provides a method for probing the mass distribution of the Universe, without any dependence on luminous tracers or physical assumptions such as hydrostatic equilibrium and virialisation. If in the early years \citep{ein01, edi01, zwi01} the discussion was essentially theoretical, in the recent years the great quantity of observed gravitational lenses changed this situation. Moreover, it was found that this phenomenon can provide precise information about the geometry of the universe and the present accelerated expansion process.
\par
In general the methods using the gravitational lensing can be classified in three cases \citep{silvia}. In the first case the time differences for images and the subsequent lens mapping of the paths followed by the light is made using the Fermat`s principle. The expression for the geometric time delay is
\begin{equation}
c\Delta t = (1 + z_l)\frac{D_{ol}~D_{os}}{2D_{ls}}(\theta - \beta)^2\quad,
\end{equation}
where, $c$ is the speed of light, $z_l$ is the cosmological redshift of the lens, $D_{ol}$ is the angular distance between the observer and the lens, $D_{os}$ is the angular distance between the observer and the source, $D_{ls}$ is the angular distance between the lens and the source, $\theta$ is the unobserved angular position of the source and $\beta$ is the observed position of the source image. In a background metric with $k = 0$, as considered here, we have $D_{os} = D_{ol} + D_{ls}$. The second case studied in gravitational lensing are the one related with the deflection suffered by light rays passing close to a massive body, considered here as a point-like deflector. This method is called the bending angle or the deflection angle wich is defined as the difference between the initial and final light ray direction and is given by
\begin{equation}
\label{alp1}
\alpha = \frac{4GM}{r_m~c^2}\quad,
\end{equation}
where $G$ is Newton's gravitational constant, $M$ is the mass of the spherical body and $r_m$ is the minimal distance between the light ray and the body of mass $M$. The more recent studies of gravitational lenses are related to statistical gravitational lensing. The general motivation for the statistical treatment of the gravitational lenses is to obtain a detailed knowledge about the matter content of the Universe. On the other side, a statistical study about gravitational lenses can provide the probability that a given background source have a multiple images under some special conditions given the expected number of lenses within an angular distance of the line of sight, which is called the optical depth.
\par
The purpose of the present work is to verify the behavior of some cosmological models from the point of view of the optical depth employing the framework developed by \cite{turner01} and \cite{tur01}. The goal here is to show a qualitative analysis of this phenomenon by comparing some cosmological models for the dark Universe. The outline of the paper is as follows: We describe the mathematical structure of statistics of gravitational lenses used in this study in terms of optical depth (probability of a lensing event occurs) in Section 2; section 3 deals with cosmological models for the dark sector and for this we chose the CDM model, the $\Lambda$CDM model, the Bose-Einstein condensate dark matter model, the Chaplygin gas model, the viscous fluid cosmological model and the holographic dark energy model. Finally we discuss the results obtained in section 4.

%%%%%%%%%%%%%%%%%%%%%%%%%%%%%%
\section{The statistics of gravitational lenses}
%%%%%%%%%%%%%%%%%%%%%%%%%%%%%%

We need a cosmological scenario to develop a statistical study of gravitational lenses. In this work our description is made with the line element
\begin{equation}
\label{metric}
ds^2 = dt^2 - a^2 (t)\biggl[d\chi^2 + f^2(\chi)(d\theta^2 + \sin^2\theta~d\phi^2)\biggr]\quad,
\end{equation}
where $t$ is the proper time coordinate, $a(t)$ is the scale factor of the Universe, $\chi$, $\theta$ and  $\phi$ are the comoving angular coordinates and $f(\chi)$ is a trigonometric, linear, or hyperbolic function of $\chi$, depending on whether the curvature $k$ is positive, zero or negative, that is the condition required by the homogeneity of the space-time. From now on we shall use natural units with $c = 1$. 
\par
In this work, two kind of distances will be fundamental for the description of the relation of the cosmological models with the optical depth
\begin{enumerate}
\item the angular diameter distance $D_{ang}$, defined  as the ratio between the proper diameter of an object at $z_2$ and the observed angular diameter of the source $D_{ang}\equiv\frac{D}{\theta} = a(z_1)~r_s$,
\item the luminosity distance $D_{lum}(z_1,z_2)$, defined by the relation in flat spacetime between the luminosity $\mathcal{L}$ of an object at $z_2$ and the flux $\mathcal{S}$ received by an observer at $z_1$, $D_{lum}\equiv\mathcal{L}/4\pi\mathcal{S} $. For a source emitting lights at time $t_1$ located in $r = r_1$ and a detector at $r= 0$ detecting the light at $t= t_0$ we have
\begin{equation}
D_{lum} = \biggl(\frac{a(z_1)}{a(z_2)}\biggr)^2 D_{ang}(z_1, z_2)\quad.
\end{equation}
\end{enumerate}
\par
An important result obtained through the study of the statistical lensing is the information about the cosmological parameters and its constraints. Suppose that the galactic lensing can be represented by a simple model known as the singular isothermal sphere (SIS). This model is, in general, consistent with the various data of the gravitational lensing, galactic dynamics and the X-ray emissions of the elliptical galaxies \citep{ofek00, ofek01, fabian01, rix01, rix02, treu01, koop02, koop00, koop01, auger00} and it reproduces very well the flat rotation curves of galaxies. More accurate and detailed values of parameters of halo density profiles related with the several dark matter models are obtained, for example, by using standard NFW profile \citep{NFW00, NFW01} whose three-dimensional density is
\begin{equation}
\rho(r) = \frac{\rho_0}{(r/r_s)(1 + r/r_s)^2} \quad,
\end{equation}
where $\rho_0$ is a constant normalization and $r_s$ is the scale radius. Another way is to use N-body simulations. We point out that both studies are not the goal of this paper. The fact is that the discussion about what is the best fit for the halos of dark matter is still open \citep{treu02, treu03} and we intend to make the comparative studies about these models in future work.
\par
In relation to the SIS model we know that the constant bend angle, equation (\ref{alp1}), which deflects the light ray due to a source as a galaxy in its own rest frame, is given by 
\begin{equation}
\label{alp}
\alpha = 4\pi\biggl(\frac{\sigma_v}{c}\biggr)^2\quad,
\end{equation}
where $\sigma_v = v_c/\sqrt 2$ is the velocity dispersion associated with the circular velocity $v_c$ of the galaxy. The velocity dispersion is the statistical dispersion of the velocities about the mean velocity for a group of objects, such as an open cluster, globular cluster, galaxy, galaxy cluster, or supercluster. By measuring the radial velocities of its members, the velocity dispersion of a cluster can be estimated and used to derive the cluster's mass from the virial theorem. The empirical correlation between the intrinsic luminosity (proportional to the stellar mass) of a spiral galaxy and how fast they are rotating is called the Tully-Fisher relation \citep{TF}. Specifically the relation is $L\propto v^4$. The Faber-Jackson relation \citep{FJ} is an early empirical power-law relation between the luminosity and the central stellar velocity dispersion  of elliptical galaxies, $L\propto\sigma_v^{\gamma}$, where $\gamma$ depends on the range of galaxy luminosities that is fitted.
\par
In addition to the relationship between luminosity and velocity dispersion there are other features of interest in the study of gravitational lenses. The distribution of the luminosity of galaxies is well approximated by the Schechter luminosity function \citep{sche76}
\begin{equation}
\phi(L) dL = \phi_*\biggl(\frac{L}{L_*}\biggr)^{\alpha} e^{-L/L_*} d\biggl(\frac{L}{L_*}\biggr)\quad
\end{equation}
where $L_*$ corresponds to an absolute $B$ magnitude of $-20.8~\mbox{mag}$, the index $\alpha = -1.25$ and the normalization constant $\phi_*$ is fixed in such a way that the mean luminosity-density of all galaxies is ${\mathcal{L}} =2\times 10^8~L_{\odot}/\mbox{Mpc}^3$ in the same band, with $L_{\odot}$ being the solar luminosity. The mass of dark matter halos considered in the general study of large-scale structures, as well as gravitational lensing as done here, is called the mass function $n(M)$, defined by $dN = n(M)dM$, where $dN$ is the number of structures per unit volume with mass between $M$ and $M + dM$. The Press-Schechter formalism \citep{sche74} predicts that the number of objects with mass between $M$ and $M + dM$ is
\begin{equation}
n(M)dM = -\sqrt{\frac{2}{\pi}}\frac{d\sigma_M}{dM}\frac{\rho_0\delta_c}{M\sigma^2_M}\mbox{exp}\biggl(-\frac{\delta_c^2}{2\sigma^2_M}\biggr)dM\quad,
\end{equation}
where $\sigma_M$ is the variance of the density field filtered on a scale $R$ enclosing a mass $M$, $\rho_0$ is the uniform background density and $\delta(\vec x)$ is the density fluctuation field. This mass function, with $\delta_c = 1.686$, gives us the number density of collapsed objects per unit mass. So, with the Press-Schechter mass function we can modelling the lenses as a population of dark matter halos. Here we assume the
mass density function as constant, $n(M) = n_0$, without limiting the generality of the study done.
\par
When $\alpha$, equation (\ref{alp}), is constant, the Einstein angle of a beam passing at any radius through the SIS is 
\begin{equation}
\theta_{E} = 4\pi\biggl(\frac{\sigma_v}{c}\biggr)^2\frac{D_{ls}}{D_{os}} \quad.
\end{equation}
\par
The expected number of lenses within an angle $\theta_E$ of the line of sight, with $d\tau$ representing the differential probability that a given background source have multiple images \citep{turner01}, is given by 
\begin{equation}
d\tau = n_0\pi\alpha^2 a_0^3\biggl(\frac{f(\chi_l)[f(\chi_s) - f(\chi_l)]}{f(\chi_s)}\biggr)^2~df(\chi_l)\quad,
\end{equation}
where the comoving galaxy density measured today $n_0 = n(z = 0)$ considered here is a constant, $a_0 = a(t_0)$ is the scale factor of the Universe at the present epoch, $f(\chi)$ is, as the line element (\ref{metric}), a trigonometric, linear, or hyperbolic function of $\chi$, depending on whether the curvature $k$ is positive, zero or negative and $l$ and $s$ correspond to the lenses and the sources, respectively.
\par
The recent results from the measurements of CMB spectrum provide $100\Omega_k = -4.2^{+4.3}_{-4.8}$ (95\%; Planck $+$ WMAP polarization low-multipole likelihood $+$ high-resolution CMB data) \citep{planck}. Hence, we can fix $k = 0$ without an oversimplification of our cosmological models. Moreover, an inflationary phase in the primordial Universe predicts (except for very special cases) $k = 0$. Since for the photon we have $ds^2 = 0 = dt^2 - a^2(t)~dr^2$ and also as we work in a flat Universe where $f(\chi) = r_s$, the comoving coordinate distance is given by
\begin{equation}
\label{r1}
r_s = \int_{0}^{r_s}~dr = \int_{t_{em}}^{t_{obs}}~\frac{dt}{a(t)}\quad.
\end{equation}
So, using the SIS model and considering $k = 0$ it is straightforward to verify that the total optical depth is given by
\begin{equation}
\label{tau}
\tau = \frac{{\cal F}}{30}r_s^3\quad,
\end{equation}
where ${\cal F}$ is a dimensionless parameter given by
\begin{equation}
{\cal F}\equiv 16\pi^3 n_0 a_0^3\biggl(\frac{\sigma_v}{c}\biggr)^4\quad.
\end{equation}
\par
The equation (\ref{tau}) gives the probability to occur the lensing phenomenon and the information about the cosmological model appears in the $r_s$ comoving distance to the source. We shall see below how to obtain the expression for  $r_s$ in some cosmological models.

%%%%%%%%%%%%%%%%%%%%%%%%%%%%%%%%%%%%%%%%%%%%%%%%%%
\section{Description of the cosmological models}
%%%%%%%%%%%%%%%%%%%%%%%%%%%%%%%%%%%%%%%%%%%%%%%%%%

Using the metric (\ref{metric}) and the Einstein's field equations
\begin{equation}
R^{\mu}_{\nu} - \frac{1}{2}Rg^{\mu}_{\nu} = 8\pi GT^{\mu}_{\nu}\quad,
\end{equation}
we obtain the Friedmann equation
\begin{equation}
\label{hubble}
H^2 = \frac{8\pi G}{3}\rho_{tot} = \frac{8\pi G}{3}\sum_i\rho_{i}\quad,
\end{equation}
where $H\equiv \dot a/a$ is the Hubble parameter and $\rho_{tot} = \sum_i\rho_{i}$ is the total matter density of the Universe.
\par
The Friedmann equation (\ref{hubble}) can be rewritten as
\begin{equation}
h^2(t) = \frac{H^2 (t)}{H_0^{2}} = \Omega_{tot} = \sum_i\Omega_{i}\quad,
\end{equation}
where $h(t)$ is the normalized Hubble parameter,  $H_0$ is the Hubble parameter today (we adopt $H_0 = 70 \textrm{km}/\textrm{s}/\textrm{Mpc}$), $\Omega_i = \rho_i/\rho_{cr}$, with $\rho_i$ denoting a matter component of the Universe, $\rho_{cr} = 3H_o^2/8\pi G$ is the critical density and the scale factor today was normalized to unity, $a(t_0) = 1$. In this way, the expression above has become a dimensionless equation. So, in terms of the redshift $z$, we have the definitions
\begin{eqnarray}
z &=& \frac{1}{a(t)} - 1\nonumber\quad,\\
H = \frac{\dot a}{a} &=& - \frac{1}{1 + z}\frac{dz}{dt}\quad,
\end{eqnarray}
and the equation (\ref{r1}) for the comoving distance is given by
\begin{equation}
\label{erre}
r_s(z_{em}, z_{obs}) = H_0^{-1}\int^{z_{em}}_{z_{obs}}~\frac{1}{h(z)}~dz\quad,
\end{equation}
where $h(z)$ is different to every cosmological model used here, as we shall see below.

%%%%%%%%%%%%%%%%%%%%%%%%%%%%%%%%%%%%%%%%
\subsection{Cold Dark Matter Model}
%%%%%%%%%%%%%%%%%%%%%%%%%%%%%%%%%%%%%%%%

By using accurate measurements of the Cosmic Microwave Background fluctuations, WMAP  determined that the Universe is flat \citep{wmap}. It follows that the mean energy density in the Universe is equal to the critical density $\rho_c$. From this total density, we now know that:
\begin{enumerate}
\item $\sim 5\%$ is made up of atoms, but this amount of baryonic mass do not explain the rotational curves of spiral galaxies and the structure formation on large scales that is observed today;
\item $\sim 23\%$ is composed of one or more species of particles that interact very weakly with ordinary matter and are modeled as pressureless and non-relativistic particles, called cold cark matter (CDM).
\item $\sim 72\%$ is made up of something that we called dark energy (DE), which has a repulsive gravitational effect and has the needed amount to explain both the flatness of the universe and the current observed accelerated expansion.
\end{enumerate}
\par
The simplest model of the Universe, which will be called CDM model, is not realistic from the observation point of view but it is an interesting toy model. It is composed only of pressureless matter, with an equation of state given by $p=0$, and radiation, with $p = \rho/3$. In this case the probability of one event of gravitational lensing occur is obtained by direct integration of equation (\ref{tau}), with help of the expression (\ref{erre}) and with the normalized Hubble parameter given by
\begin{equation}
h(z) = [\Omega_m~(z + 1)^3 + \Omega_r~(z + 1)^4]^{1/2}\quad,
\end{equation}
where $\Omega_m$ is the dark matter density parameter (baryonic and non-baryonic) and $\Omega_r$ is the radiation density parameter.
\par
\begin{figure}[tb] 
\includegraphics[width=\columnwidth]{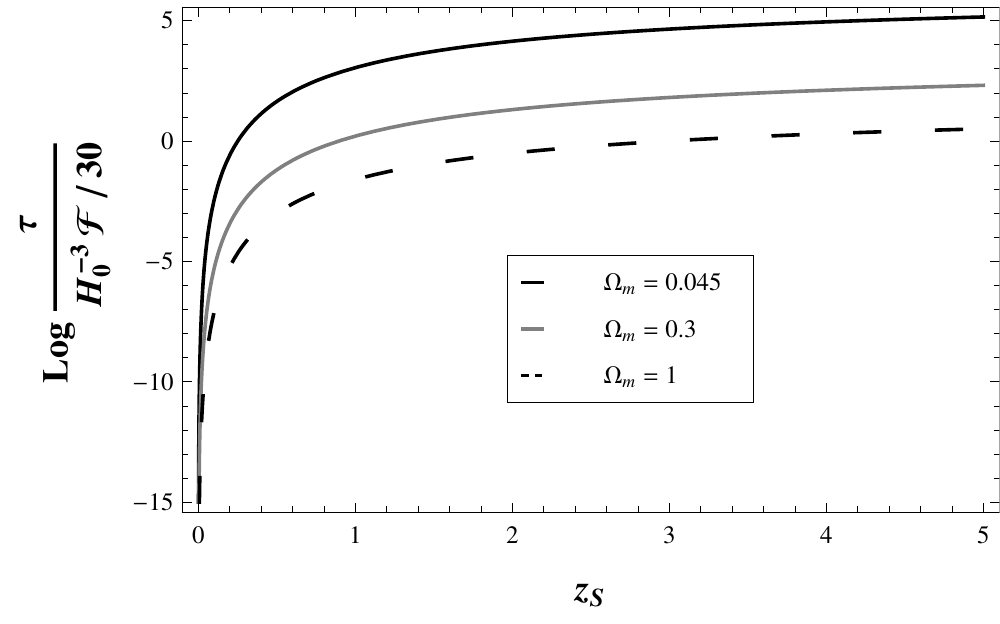}
\caption{Optical depth behavior as a function of the source redshift $z_S$ in the CDM model for different values of $\Omega_m$ and with $\Omega_r = 10^{-4}$.}
\label{fig_cdm1}
 \end{figure}

The curve for this model can be viewed in Figure \ref{fig_cdm1} with some values of the density parameter of the pressureless matter $\Omega_m$ and with the value of the radiation $\Omega_r$ fixed. For the estimation of the radiative contribution, neutrino and photon components are taken into account. For the pressureless fluid, baryonic and weakly interacting massive particles (WIMPS), which can be components of the cold dark matter, are considered.
\par
In Figure \ref{fig_cdm1} we see that for $\Omega_m = 0.045$ the optical depth is substantially larger as compared with the observational estimation, $\Omega_m = 0.3$, and when the dark matter dominates the mass content of the Universe, $\Omega_m = 1$.

%%%%%%%%%%%%%%%%%%%%%%%%%%%%%%%%%%%%%%%%
\subsection{$\Lambda$CDM}
%%%%%%%%%%%%%%%%%%%%%%%%%%%%%%%%%%%%%%%%

A cosmological model with a positive cosmological constant ($\Lambda > 0$), formed by an exotic form of energy with an equation of state $p = -\rho$ (the same equation of state for the vacuum) is called $\Lambda$CDM model. The cosmological constant fluid has a density parameter of about $\Omega_{\Lambda}\equiv\Lambda/3H^2\approx 0.7$ and the pressureless non-baryonic dark matter, which does not couple with radiation, have a density of about $\Omega_m\approx 0.3$. This cosmological model, also called concordance model, in general solve the problems of the accelerated expansion of the Universe and the rotation curves of spiral galaxies. To a flat universe ($\Omega_{k}=0$) described by this model, the normalized Hubble parameter is given by
\begin{equation}
h(z) = [\Omega_m~(z + 1)^3 + \Omega_{\Lambda}]^{1/2}\quad.
\end{equation}
\par
\begin{figure}[tb] 
\includegraphics[width=\columnwidth]{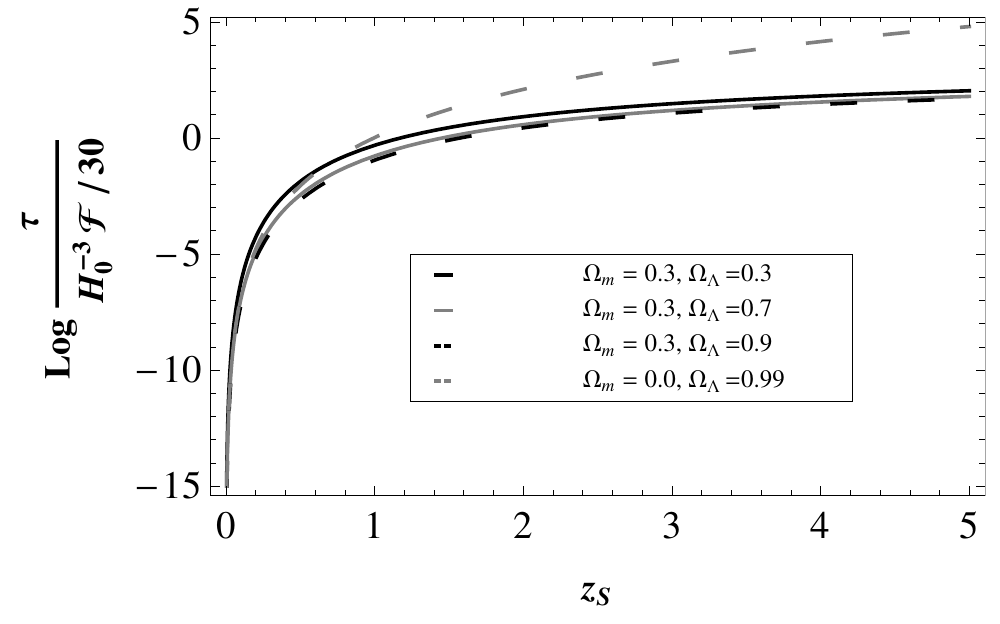}
\caption{Optical depth behavior as a function of the source redshift $z_S$ in the $\Lambda$CDM model for different values of $\Omega_m$ and $\Omega_{\Lambda}$.}
\label{fig_lcdm1}
 \end{figure}

The optical depth for this model is represented in Figure \ref{fig_lcdm1} where it can be observed almost no differences among the cases $(\Omega_m = 0.3 ,~ \Omega_{\Lambda} = 0.3)$, $(\Omega_m = 0.3 ,~ \Omega_{\Lambda} = 0.7)$ and $(\Omega_m = 0.3 ,~ \Omega_{\Lambda} = 0.9)$. When there is only the cosmological constant, the optical depth $\tau$ become larger than the other cases. As the concordance model is more acceptable from the standpoint of cosmological observations, in the case of statistical lensing this model cannot be confirmed yet by the current observational data.

%%%%%%%%%%%%%%%%%%%%%%%%%%%%%%%%%%%%%%%%
\subsection{Bose-Einstein Condensate Dark Matter Model}
%%%%%%%%%%%%%%%%%%%%%%%%%%%%%%%%%%%%%%%% 

The fundamental nature of dark matter is still unknown. The existence of dark matter particles, such as WIMPs, is one of the possible hypotheses considered in order to describe the observable behavior of the CDM. If these particles are spin-$0$ bosons a Bose-Einstein condensation \citep{bose01} can take place during the history of the Universe once the temperature of the dark matter gas is smaller than the critical temperature. In this model dark matter is described as a non-relativistic gravitational condensate with a polytropic equation of state \citep{bose02, bose03}.
\par
\begin{figure}[tb!] 
\includegraphics[width=\columnwidth]{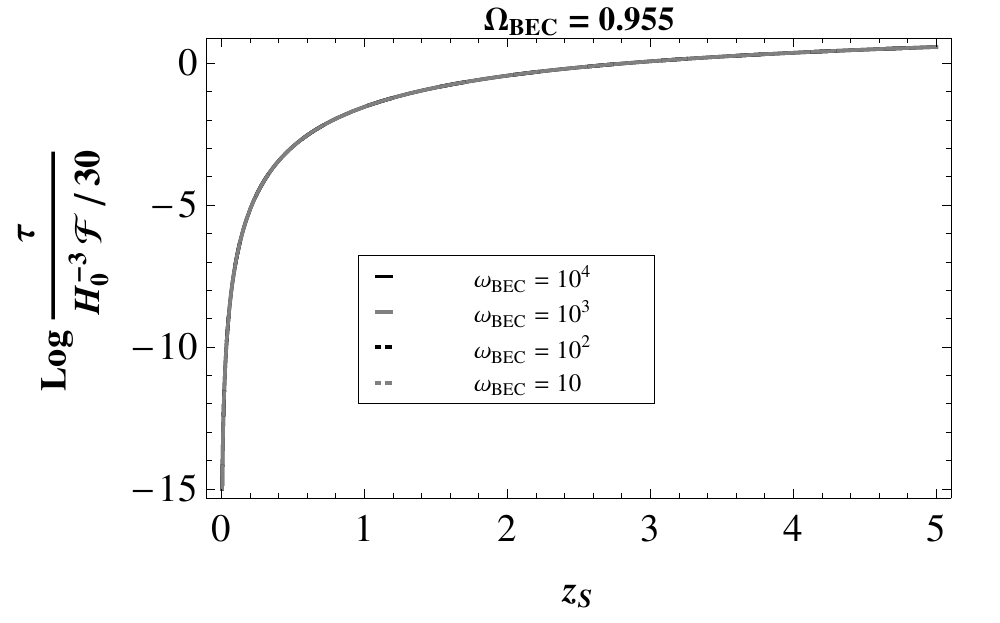}
\includegraphics[width=\columnwidth]{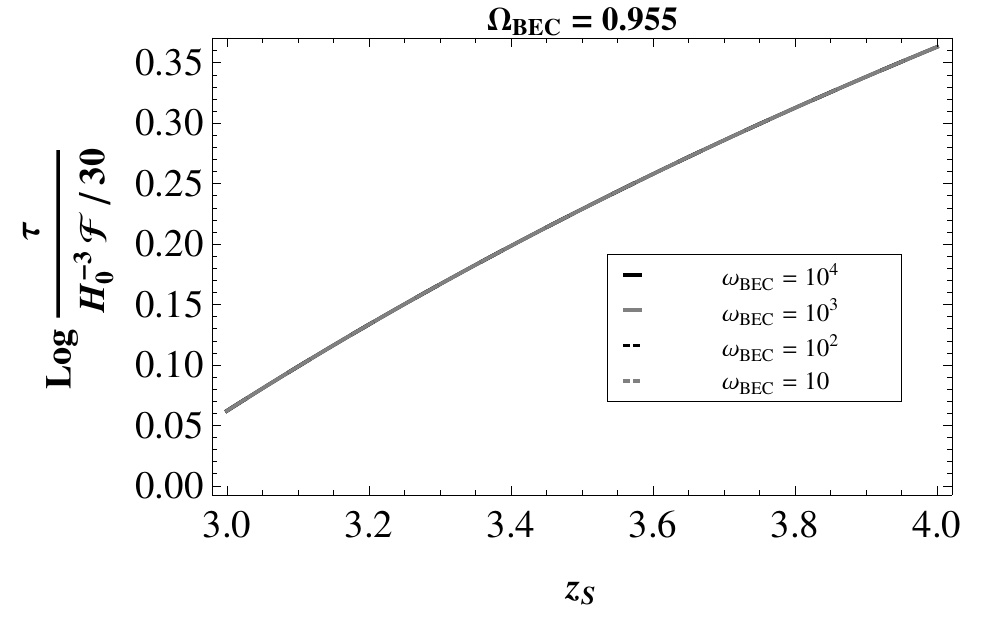}
\includegraphics[width=\columnwidth]{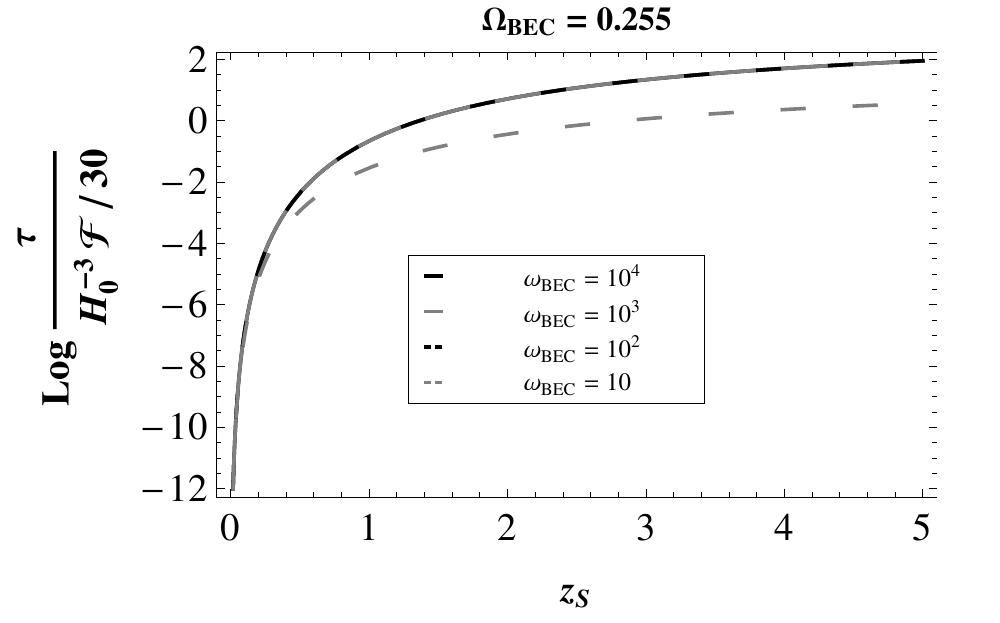}
\caption{Optical depth behavior as a function of the source redshift $z_S$ in the Bose-Einstein condensate dark matter model for different values of $\omega_\textrm{BEC}$ without (upper figure) and with cosmological constant (lower figure). To allow a closer inspection, the middle figure we draw the same figure above, but on a reduced scale. The overlap of the curves remain.}
\label{fig_bec1}
\end{figure}
\begin{figure}[tb] 
\includegraphics[width=\columnwidth]{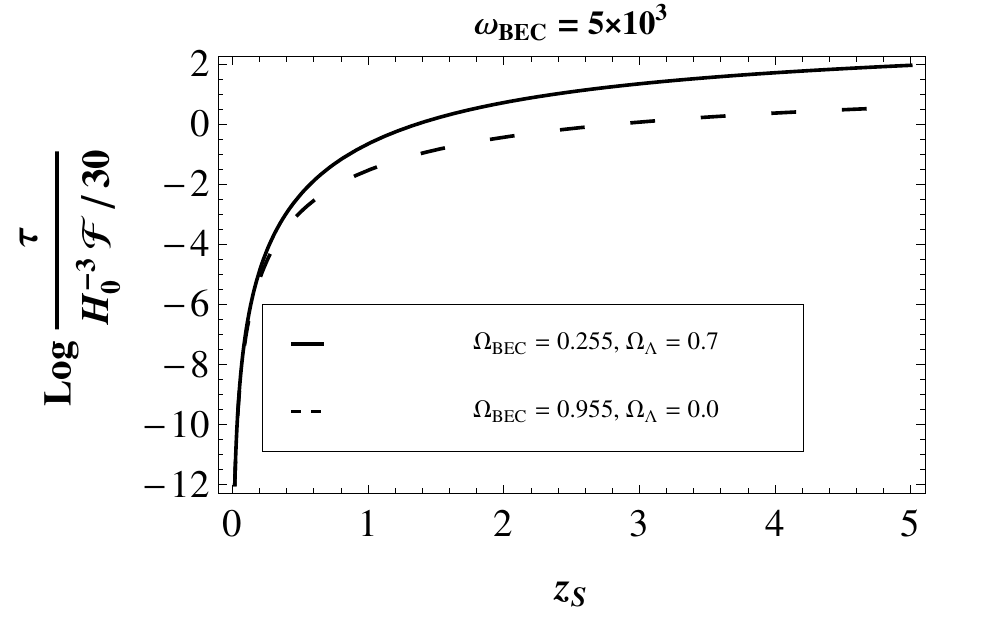}
\caption{Comparing the optical depth as a function of the source redshift $z_S$ in the BEC model for a fixed value of $\omega_\textrm{BEC}$ in Universes filled by baryons, BEC dark matter and cosmological constant and only by baryons and BEC dark matter. The critical density is $\rho_{crt} = 10^{-27}$ Kg/m$^3$.}
\label{fig_bec2}
\end{figure}

After the condensation process all dark matter is in the form of Bose-Einstein condensate (BEC) and the equation of state is
\begin{eqnarray}
P = \omega_{\textrm{BEC}} \rho^2 \quad, \\
\omega_{\textrm{BEC}} = \frac{2 \pi \hbar^2 l_a}{m^3} \quad,
\end{eqnarray}
where $l_a$ is the scattering length and $m$ is the mass of the dark matter particles. Using the energy-density conservation equation we find that
\begin{equation}
   \rho_{\textrm{BEC}}=\frac{\rho_0\left(1+z\right)^3}{\left(1+\omega_{\textrm{BEC}}\rho_0\right)-\omega_{\textrm{BEC}}\rho_0\left(1+z\right)^3}\quad,
\end{equation}
and the reduced Hubble parameter for a Universe filled with baryonic matter, BEC and cosmological constant is

\begin{eqnarray}\nonumber
&&h(z) = \biggl[\Omega_{\Lambda} + \biggl(\Omega_m + \\
&& \frac{\Omega_{\textrm{BEC}}}{\left(1+\omega_{\textrm{BEC}}\rho_0\right)-\omega_{\textrm{BEC}}\rho_0\left(1+z\right)^3}
\biggr)(1 + z)^3\biggr]^{1/2}\quad, \end{eqnarray}
where $\rho_{\textrm{crt}}$ is the critical density of the Universe today. Here we will assume the typical values $l_a = 10^{-12}~\textrm{m}$ and $m=10^{-36}~\textrm{Kg}$ such as $\omega_{\textrm{BEC}} \approx 10^3$ and $\rho_{\textrm{crt}}=10^{-32}~\textrm{Kg}/\textrm{m}^3$. In Figure \ref{fig_bec1} we show the curve of optical depth for the BEC dark matter model with different values of $\omega_{BEC}$ with and without cosmological constant. When the cosmological constant is absent, {\it i.e.}, the total quantity of matter is made of the condensate dark matter (upper figure), the optical depth is insensitive to the chosen values of the parameter $\omega_{BEC}$. This is more clear in the middle figure where the same values of the cosmological parameters of the above figure are used. The overlap of the curves remains indicating that this result is a consequence of the little influence of the variation of the parameters in the calculation of the optical depth. However, when the cosmological constant is included (lower figure) we see that for low values of $\omega_{BEC}$ this cosmological model produces $\tau$ with a lower probability of finding gravitational lenses, while for large values of the parameter $\omega_{BEC}$ no difference is found. The Figure \ref{fig_bec2} shows the comparison of the BEC dark matter model for a fixed value of $\omega_{BEC}$ with baryons, dark matter and cosmological constant (solid black line) and only baryons and dark matter (dashed line). We can see that the case with cosmological constant produces a larger probability of finding gravitational lenses compared with the case without cosmological constant.

%%%%%%%%%%%%%%%%%%%%%%%%%%%%%%%%%%%%%%%%
\subsection{Generalized Chaplygin Gas Model}
%%%%%%%%%%%%%%%%%%%%%%%%%%%%%%%%%%%%%%%%

The cosmological data indicate that there are two basic dark matter-energy components in the Universe and we still do not know their nature. But, by definition, we can assume that the dark matter and the dark energy are two forms of  a one single fluid. This scenario is called Unified Dark Matter Energy (UDME)  or quartessence, \textit{i.e.}, models in which these two dark components are seen as different manifestations of a single fluid \citep{cha01}. Among the theoretical proposals of this scenarios, an interesting attempt of unification is called generalized Chaplygin gas model (GCG) \citep{cha02}. This exotic fluid has an equation of state given by $p = -A\rho^{-\alpha}$ where $A$ is a positive constant, $\alpha$ is a constant in the range $0 < \alpha \leq 1$ with $\alpha = 1$ being the ordinary Chaplygin gas (CG). For this situation the solution of the continuity equation and the normalized Hubble parameter are written as
\begin{eqnarray}
&\rho_{Ch}& = \rho_0 \times\nonumber\\
&&\biggl(A_{Ch} + (1-A_{Ch})(1 + z)^{3(1 + \alpha)}\biggr)^{\frac{1}{1 + \alpha}},\\
\label{hcgm}
&h(z)& = \nonumber\\
&&\left[\Omega_{Ch}\left(A_{Ch} + (1-A_{Ch})\right)(1+z)^{3(1 + \alpha)} \right]^{\frac{1}{2(1 + \alpha)}}\quad .
\end{eqnarray}
where $A_{Ch} = A/\rho_{0}^{1 + \alpha}$ is a quantity related to the sound speed of the Chaplygin fluid today, with $v_s^2 = \alpha A_{Ch}$. Using the expression (\ref{hcgm}) in the equations (\ref{erre}) and (\ref{tau}) we obtain the optical depth for this model. Currently the CG model has been adopted as a candidate for the DE only \citep{FA02.1}.
\par
If we include a contribution from dark matter and pressureless matter, since this is not accounted by the generalized equation of state of the Chaplygin gas, we will have a new ensemble of models whose normalized Hubble parameter is given by

\begin{eqnarray}
h(z)&=& \biggl[ \Omega_{Ch}\left(A_{Ch}+(1-A_{Ch})\right)(1+z)^{3(1 + \alpha)}+\nonumber\\
& &\Omega_{m}(1+z)^3\biggr]^{\frac{1}{2(1 + \alpha)}}\quad.
\end{eqnarray}

\begin{figure}[tb] 
\includegraphics[width=\columnwidth]{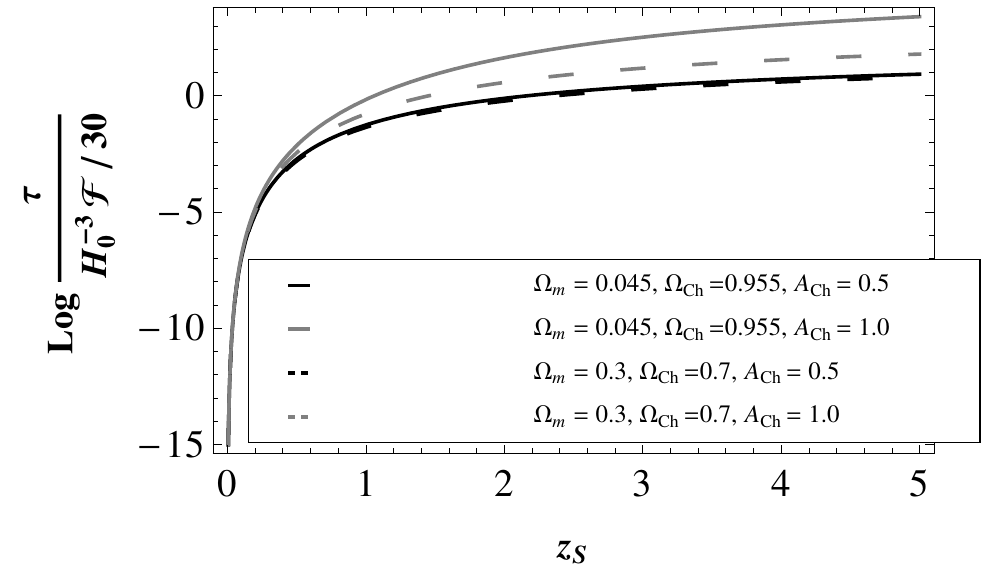}
\includegraphics[width=\columnwidth]{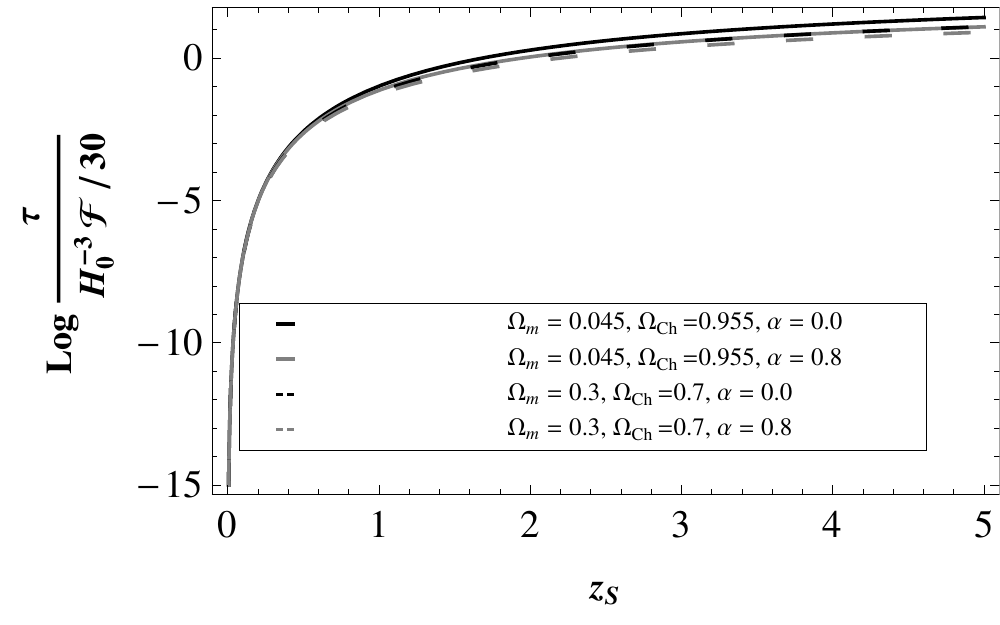}
\caption{Optical depth as a function of the source redshift $z_S$ in the Chaplygin gas model, $\alpha = 1$, (upper figure) and in the generalized Chaplygin gas model (lower figure), with $A_{Ch} = 0.6$.}
\label{Cgm1}
\end{figure}

In Figure \ref{Cgm1} we can observe the models CG and GCG together for different values of the theoretical parameters. The upper figure, Chaplygin gas model, shows that the large values of the parameter $A_{CH}$ and $\Omega_{CH}$ produce large probability of finding gravitational lenses. The case where $A_{Ch} = 0.5$ is insensitive to the quantity of Chaplygin gas density $\Omega_{Ch}$. The models $(\Omega_m = 0.045,~\Omega_{Ch} = 0.955)$ and $(\Omega_m = 0.3,~\Omega_{Ch} = 0.7)$ are almost indistinguishable. On the other hand, in the generalized Chaplygin gas model all cases are almost indistinguishable, irrespective of the values of the parameters $\Omega_m$, $\Omega_{Ch}$ and $\alpha$.

%%%%%%%%%%%%%%%%%%%%%%%%%%%%%%%%%%%%%%%%
\subsection{Viscous Cosmological Model}
%%%%%%%%%%%%%%%%%%%%%%%%%%%%%%%%%%%%%%%%

Other possibility to describe the dark sector is by using a viscous fluid. The equation of state in this case is \citep{vis02}
\begin{equation}
p =\beta \rho- \xi(\rho) u^\mu_{;\mu} \quad,
\end{equation} 
where $p_\beta = \beta \rho$ is the perfect fluid component and $p_{visc} = - \xi(\rho) u^\mu_{;\mu}$ is a bulk viscosity component. We consider that the bulk viscous component has a power law dependence in the energy density  according to $\xi(\rho_v)=\xi_0 \rho_v^\nu$ where $\xi_0$ is a constant. Once that $u^\mu_{;\mu} = 3 H$, we have
\begin{equation}
\label{vp}
p =\beta \rho- 3 H \xi_0 \rho^\nu \quad,
\end{equation} 
and the continuity equation with the equation of state (\ref{vp}), leads to
\begin{equation}
\rho = \rho_{v0}\left[ A_{visc} + (1-A_{visc}) (1+z)^{-r}\right]^{\frac{1}{\frac{1}{2}-\nu}},
\end{equation}
with a normalized Hubble parameter given by
\begin{equation}
\label{hvm}
h(z)=\Omega_{visc}^{1/2}\left[ A_{visc} + (1-A_{visc}) (1+z)^{-r}\right]^{\frac{1}{1-2\nu}}\quad,
\end{equation}
where the parameters are
\begin{eqnarray}
A_{visc} &=& 3\xi_0 \left(\frac{8\pi G}{3} \right)^{1/2}\frac{1}{1+\beta}\frac{1}{\rho_{vis,0}^{1/2-\nu}}\quad,\nonumber\\
r &=& 3(1+\beta)\biggl(\nu-\frac{1}{2}\biggr)\quad.
\end{eqnarray}
\par
\begin{figure}[tb]
\includegraphics[width=\columnwidth]{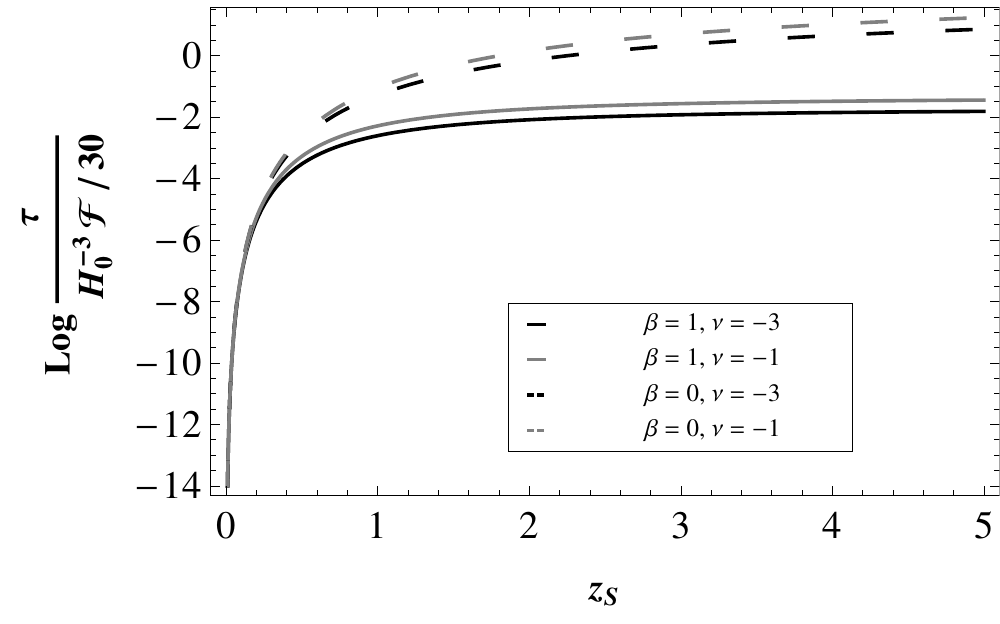}
\caption{Optical depth as a function of the source redshift $z_S$ in the viscous fluid model, with $\Omega_{visc} = 1.0$ and $A_{visc} = 0.6$.}
\label{figvisc}
\end{figure}

With the equation (\ref{hvm}) we can obtain the optical depth $\tau$ whose behavior is shown in Figure \ref{figvisc}. Here the behavior of the optical depth appears to be grouped by the value of the parameter $\beta$, associated with the barotropic fluid of the equation of state (\ref{vp}). It is insensitive to the value of the parameter $\nu$.
\par
\begin{figure}[tb]
\includegraphics[width=\columnwidth]{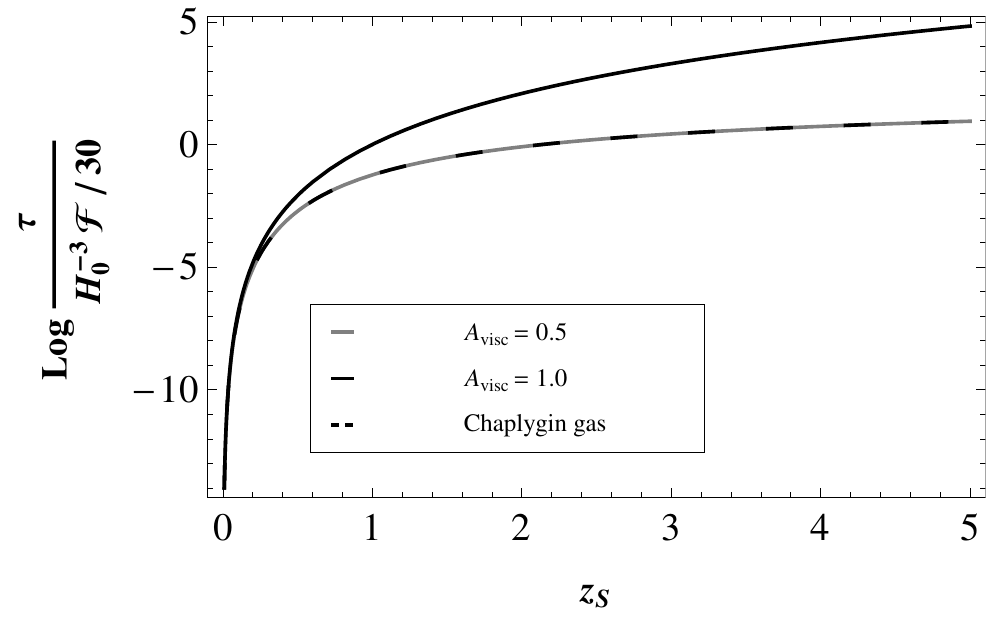}
\caption{Comparison of the optical depth as a function of the source redshift $z_S$ between the viscous fluid model, with $\Omega_{visc} = 1.0$, $\beta = 0$, and $\nu = -3/2$, and the Chaplygin gas model, with $\Omega_{Ch}=1.0$ and $A_{Ch}=0.5$.}
\label{figvisc1}
\end{figure}

The cosmological model obtained by a viscous fluid is more general than the Chaplygin gas model at background level. This is verified when we analyse the normalized Hubble parameter, equation (\ref{hvm}). If we consider the values  $\beta = 0$, $\nu = -3/2$ and $\Omega_m = 0$ the result has a similar behavior to the cosmological scenario of the Chaplygin gas model, equation (\ref{hcgm}). This result can be observed in Figure \ref{figvisc1} to the particular case where $A_{visc} = 0.5$. When $A_{visc} = 1.0$ the similar behavior between viscous model and Chaplygin gas model is lost and the probability of finding gravitational lenses is larger.
\par
\begin{figure}[tb] 
\includegraphics[width=\columnwidth]{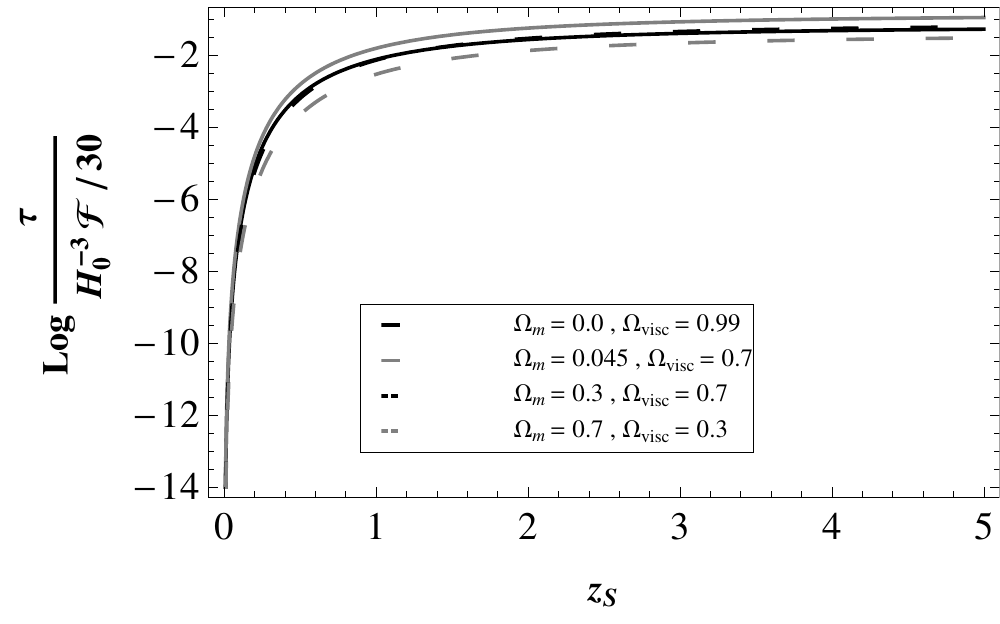}
\includegraphics[width=\columnwidth]{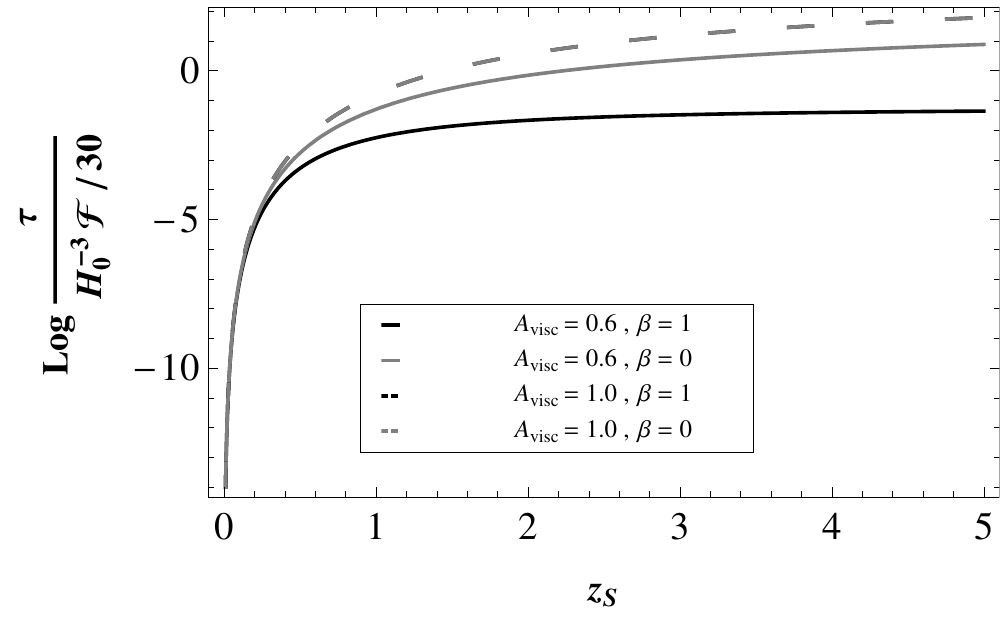}
\caption{Optical depth as a function of the source redshift $z_S$ in the Universe where there is a mixture of viscous fluid and the cold dark matter. In the upper figure $A_{visc} = 0.6$, $\beta = 1.0$ and $\nu = -1.0$ while in the lower figure $\Omega_m = 0.3$, $\Omega_{visc} = 0.7$ and $\nu = - 3/2$.}
\label{figvisc2}
\end{figure}

A Universe whose matter content is a mixture of dark matter and viscous fluid is shown in Figure \ref{figvisc2}. In this case the upper figure is similar at the upper figure of Figure \ref{figvisc1} where there is only the viscous fluid. The lower figure shows the same behavior of the model that has the parameters $(A_{visc} = 0.6, \beta = 1)$ and $(A_{visc} = 1.0, \beta = 1)$.

%%%%%%%%%%%%%%%%%%%%%%%%%%%%%%%%%%%%%%%%
\subsection{Holographic Dark Energy Model}
%%%%%%%%%%%%%%%%%%%%%%%%%%%%%%%%%%%%%%%%

The so called holographic dark energy model \citep{LI04.1,cam,HM02.1,CK99.1} represents other prototype of a unified cosmological model. In this model is established a theoretical relation between a short distance (ultraviolet cutoff) and a long distance (infrared cutoff), according to the holographic principle, where the number of degrees of freedom in a bounded system should be finite and related to the area of its boundary. Different values of the cutoffs provide different holographic dark energy models. When $L$, the volume of any space part, is identified with the Hubble radius $H^{-1}$, the resulting dark-energy density will have the value close to the observed value for the cosmological constant. For a special class of this model we can allow interaction with dark mater, doing a more realistic model. In this model, the Friedmann equations for the spatially flat case are given by
\begin{eqnarray}
H^2 &=&  \frac{8 \pi G}{3} (\rho_m + \rho_{hol})\quad,\\
\label{2F}
\dot H &=& -4\pi G(\rho_m + \rho_{hol} + p_{hol})\nonumber\\
&=& -\frac{3}{2}H^2\biggl(1 + \frac{\omega}{1 + r}\biggr)\quad,
\end{eqnarray}
where $\rho_m$ is the energy density of the pressureless matter component, $\rho_{hol}$ is the holographic dark energy component, $p_{hol}$ is the pressure associated with the holographic component, $w = p_{hol}/\rho_{hol}$ is the equation-of-state parameter and $r = \rho_m/\rho_{hol}$ is the ratio of the energies of both components.  The total energy density $\rho_t = \rho_m + \rho_{hol}$ is conserved and we can suppose that both components interact according to
\begin{equation}
\dot{\rho}_m + 3H\rho_m = Q \quad,
\end{equation}
and
\begin{equation}
\dot{\rho}_{hol} + 3H(1+w)\rho_{hol} = -Q\quad.
\label{cont_rhoH}
\end{equation}
\par
According to \citep{LI04.1} the holographic dark energy density is
\begin{equation}
\label{ded}
\rho_{hol} = \frac{3 c^2 M_p^2}{L^2}
\end{equation}
where $L$ is the infrared (IR) cutoff scale and $M_p = 1/\sqrt{8\pi G}$ is the reduced Planck mass. The numerical constant $c^2$ determines the degree of saturation of the condition $L^3 \rho_{hol} \leq M_p^2 L$, which means that the holographic dark energy density in the box of size $L$ cannot exceed the energy of a black hole of the same size.
\par
In the following we show two different choices of the cutoff scale $L$ that lead to two different holographic dark energy models and the $h(z)$ values used in the optical depth calculus.

%%%%%%%%%%%%%%%%%%%%%%%%%%%%%%%%%%%%%%%%
\subsubsection{Hubble-scale cutoff}
%%%%%%%%%%%%%%%%%%%%%%%%%%%%%%%%%%%%%%%%

In this particular choice of the cutoff we have $L = H^{-1}$ and the equation (\ref{ded}) becomes
\begin{equation}
\rho_{hol} = 3 c^2 M_p^2 H^2\quad.
\end{equation}
Differentiating the above equation, applying the equation (\ref{2F}) and using the ratio between the energies $r$ we obtain
\begin{equation}
\dot{\rho}_{hol} + 3 H (1+\omega)\rho_{hol} = \frac{3H\omega\rho_m}{1 + r}\quad.
\end{equation}
Comparing this result with the equation (\ref{cont_rhoH}) we see that the value of the interaction term $Q$ to this particular case is given by
\begin{eqnarray}
Q = -\frac{3H\omega\rho_m}{1 + r} \quad.
\end{eqnarray}
\par
\begin{figure}[tb]
\includegraphics[width=\columnwidth]{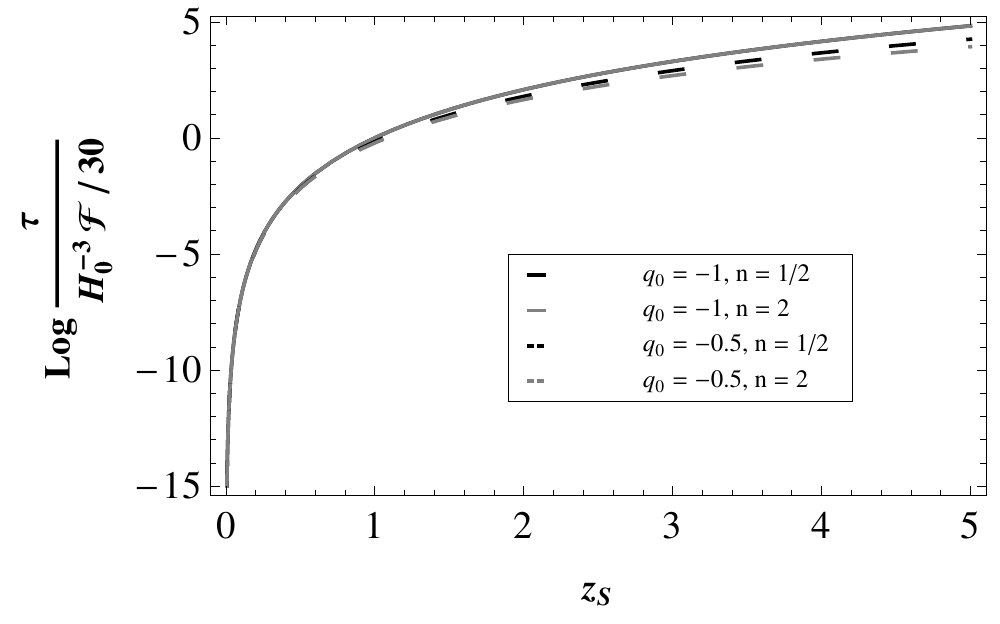}
\includegraphics[width=\columnwidth]{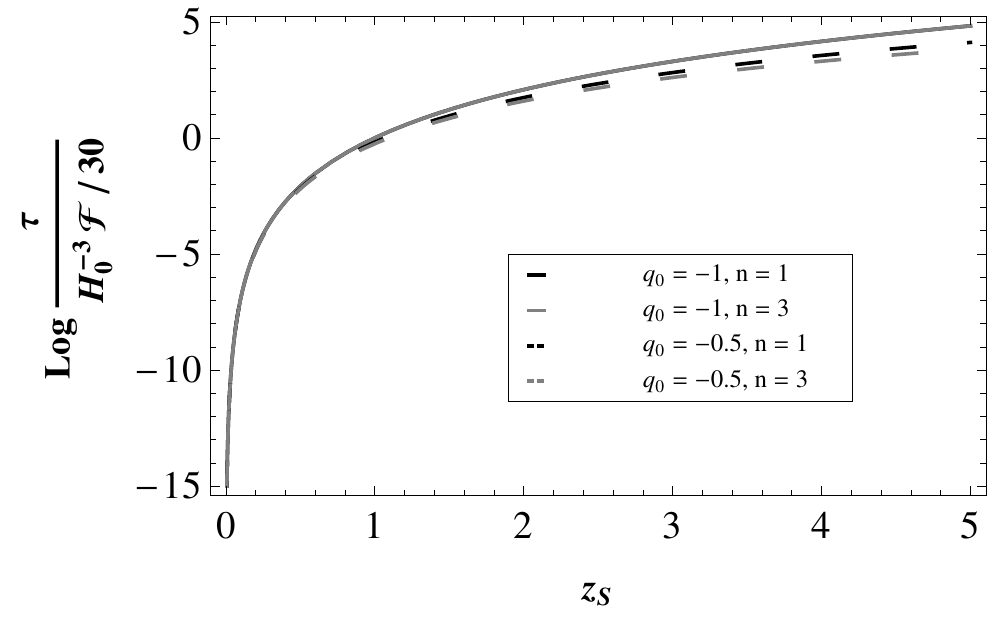}
\caption{Optical depth as a function of the source redshift $z_S$ in the holographic dark energy model with Hubble cutoff, $L = H^{-1}$, with the variation of the parameters $n$ and $q_0$ and $\Omega_{hol} = 1.0$.}
\label{fighol}
\end{figure}
\begin{figure}[tb]
\includegraphics[width=\columnwidth]{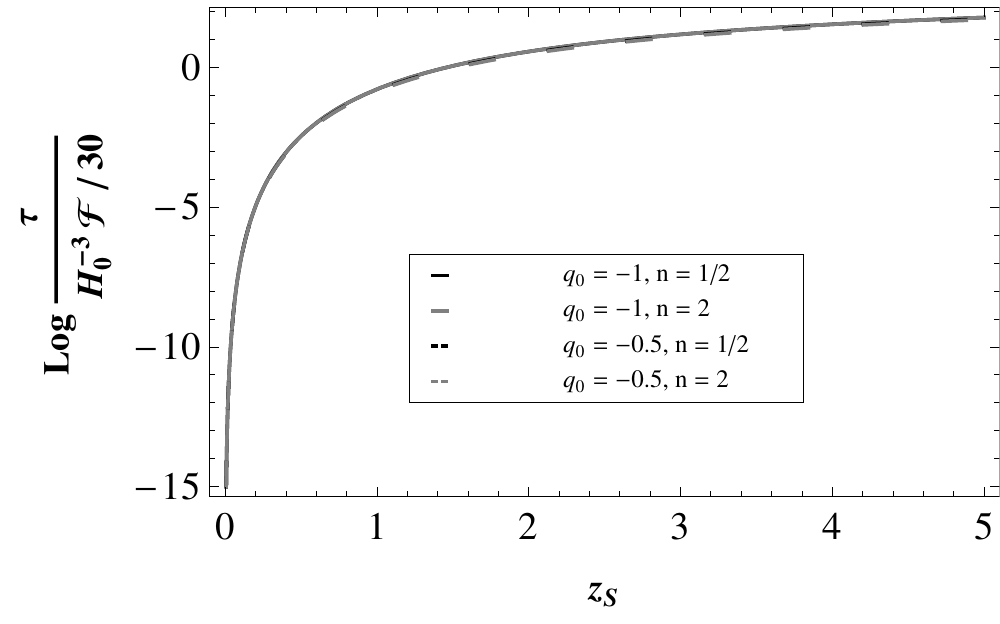}
\includegraphics[width=\columnwidth]{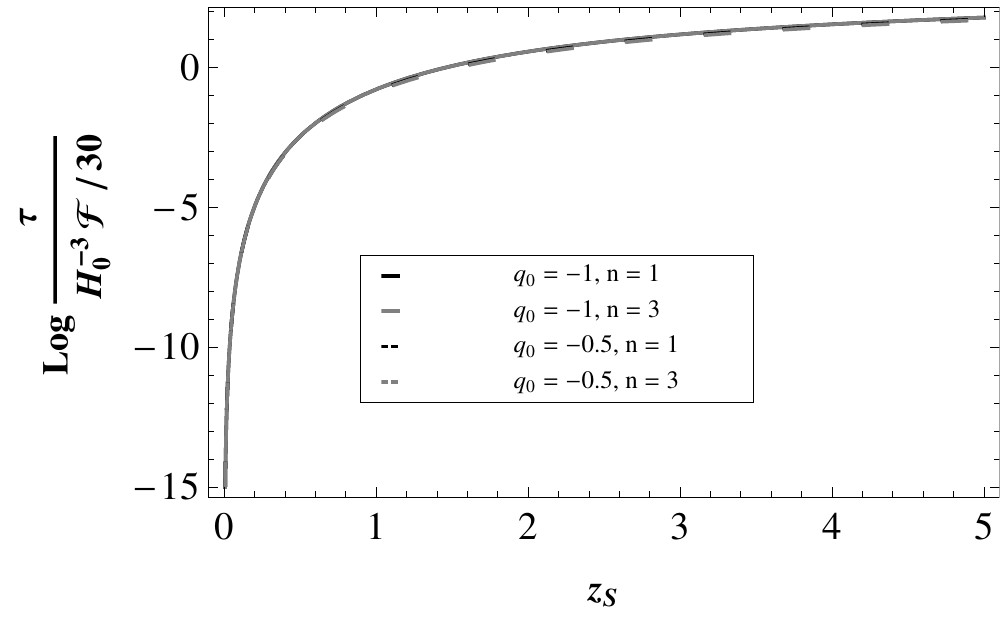}
\caption{Optical depth as a function of the source redshift $z_S$ in the holographic dark energy model in the Hubble cutoff, $L = H^{-1}$, with the variation of the parameters $n$ and $q_0$ and $\Omega_{hol} = 0.7, \Omega_m = 0.3$.}
\label{figholcdm1}
\end{figure}

We can define a new parameter, $\Gamma$ as 
\begin{equation}
\Gamma \equiv \frac{Q}{\rho_{hol}} = -\frac{3H\omega}{1 + r} r\quad,
\end{equation} 
which denotes the ratio of change of  $\rho_{hol}$ produced by the interaction. The expression $\Gamma/r$ is a freedom parameter and can be used to establish a viable cosmological model. Here, we assume that the interaction rate $\Gamma$ is proportional to a power of Hubble rate
\begin{equation}     
\frac{\Gamma}{3Hr} = \mu \left( \frac{H}{H_0}\right)^{-n}\quad .
\end{equation}
The parameter $n$ allows us to write different interactions while the quantity $\mu$ is an interaction constant and is related to the present value of the deceleration parameter $q_0$ as
\begin{equation}
\mu = \frac{1}{3} (1-2q_0) \quad.
\end{equation}
So we can write the continuity equation as
\begin{equation}
\label{cont}
\dot{\rho} + 3 H \left[ 1- \mu \left( \frac{H}{H_0}\right)^{-n}\right]\rho = 0\quad ,
\end{equation}
where $\rho = \rho_m + \rho_{hol}$ corresponds to a total density in the spatially flat background. 
The solution of equation (\ref{cont}) is equivalent to that of the generalized Chaplygin gas \citep{FA02.1}  for $n \not= 0$,
\begin{equation}
\rho = \rho_0 \left[ \mu + (1-\mu)a^{\frac{-3n}{2}}\right]^{\frac{2}{n}}
\end{equation}
so the normalized Hubble parameter, in terms of  $q_0$, is given by
\begin{equation}
\label{hub}
h(z) = \sqrt{\Omega_0}\left[\frac{1-2q_0+2(1+ q_0)(1+z)^{\frac{3n}{2}}}{3}\right]^{\frac{1}{n}}\quad .
\end{equation}
From the above expression we obtain the curves of the optical depth, for this particular case, shown in Figures \ref{fighol} and \ref{figholcdm1}. The behavior is the same in the upper and lower pictures in Figure \ref{fighol} where $\Omega_{hol} = 1.0$. On the other hand, in Figure \ref{figholcdm1}, where CDM is included, we see total similarity between the models  despite variation of parameters $q_0$ and $n$.
\par
\begin{figure}[tb]
\includegraphics[width=\columnwidth]{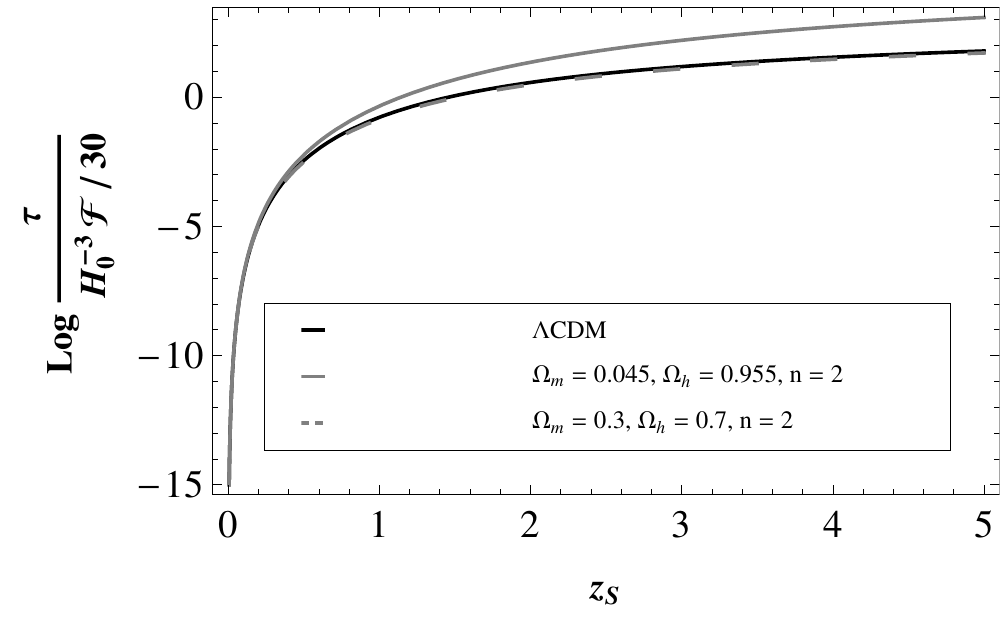}
\caption{Optical depth as a function of the source redshift $z_S$ in the holographic dark energy and pressureless matter with Hubble cutoff, $L = H^{-1}$, versus $\Lambda$CDM model. The deceleration parameter is fixed in $q_0 = - 0.5$. We see the curves overlap for $\Lambda$CDM (solid black line) and the holographic dark energy model (dashed line) with the parameters values $\Omega_m = 0.3$, $\Omega_{hol} = 0.7$ and $n = 2$.}
\label{fighollcdm}
\end{figure}

For the special case $n=2$, the expression (\ref{hub}) is similar to that for the $\Lambda$CDM model and this comparison is shown in Figure \ref{fighollcdm}. We see that the similarity occurs only when dark matter and holographic dark energy are present in the proportion of $30\%$ and $70\%$, respectively.

%%%%%%%%%%%%%%%%%%%%%%%%%%%%%%%%%%%%%%%%
\subsubsection{Future event horizon cutoff}
%%%%%%%%%%%%%%%%%%%%%%%%%%%%%%%%%%%%%%%%

Other possible cutoff can be obtained considering $L = R_E$, where
\begin{equation}
R_E(t) = a\int^{\infty}_a \frac{da'}{H(a')~a'^2}\quad.
\end{equation}
\par
So, the expression (\ref{ded}) of the holographic dark energy density can be written as
\begin{equation}
\rho_{hol} = \frac{3c^2 M^2_p}{R^2_E}\quad.
\end{equation}
Differentiating the above equation we have
\begin{equation}
\dot\rho_{hol} = -2\frac{\dot R_E}{R_E}\rho_{hol}\quad,
\end{equation}
that leading to expression of the conservation of density energy and the interaction factor $Q$
\begin{eqnarray}
\dot\rho_{hol} &+& 3H(1 + \omega)\rho_{hol} = \biggl[(1 + 3\omega)H + \frac{2}{R_E}\biggr]\rho_{hol}\quad,\nonumber\\
Q &=& - \biggl[(1 + 3\omega)H + \frac{2}{R_E}\biggr]\rho_{hol}\quad.
\end{eqnarray}
\par
Assuming a particular solution in terms of power law function for the energy density ratio as $r = \rho_m/\rho_{hol} = r_0 a^{-\epsilon}$ we can write the interaction rate parameter $\Gamma$ as
\begin{equation}
\Gamma = Hr\biggl[1 - \epsilon + \frac{2}{c}\frac{1}{\sqrt{1 + r}}\biggr]\quad,
\end{equation}
where $\epsilon < 3$ makes the coincidence problem (the fractional densities of dark matter and dark energy are about the same: $\Omega_{dm}\sim\Omega_{\Lambda}$) less severe than in the $\Lambda$CDM model, where $\epsilon = 3$. Below, the Hubble parameter is shown in general case.
\begin{eqnarray}
h(z, \epsilon) &=& (1 + z)^2\frac{\sqrt{r_0 + (1 + z)^{-\epsilon}}}{\sqrt{r_0 + 1}}\times\nonumber\\
&\times&\biggl[\frac{\sqrt{r_0 + (1 + z)^{-\epsilon}} + \sqrt{(1 + z)^{-\epsilon}}}{\sqrt{r_0 + 1} + 1}\biggr]^{\frac{2}{\epsilon c}}\quad,
\end{eqnarray}
\par
\begin{figure}[tb]
\includegraphics[width=\columnwidth]{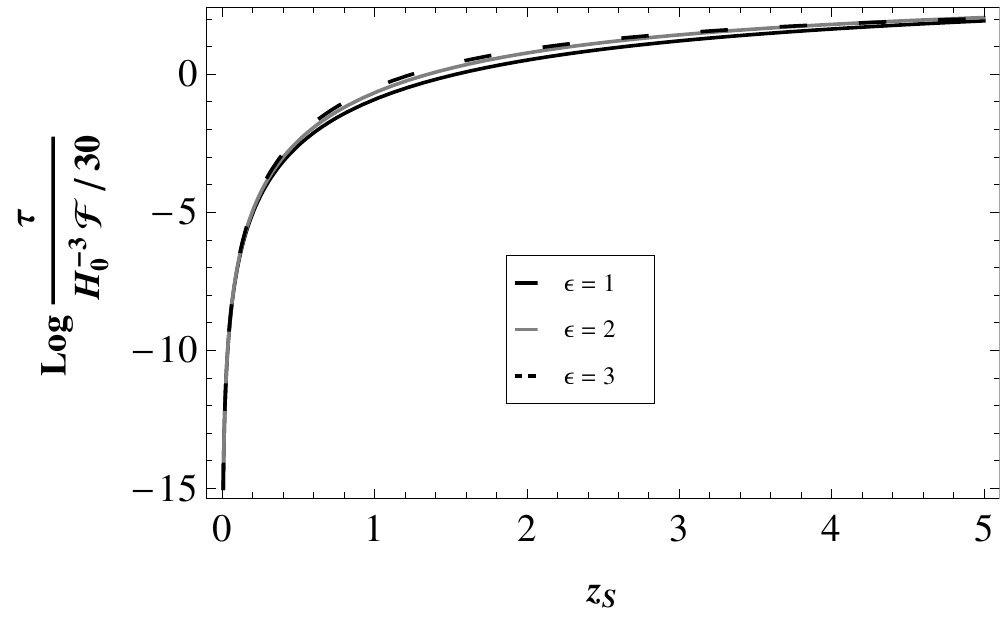}
\includegraphics[width=\columnwidth]{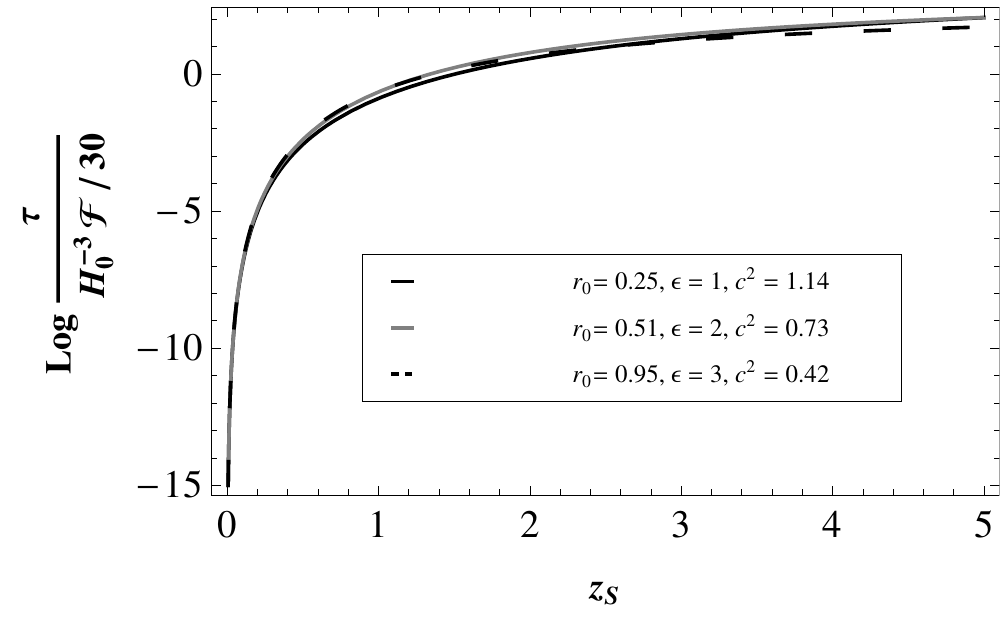}
\caption{Optical depth as a function of the source redshift $z_S$ in the holographic dark energy model with  future event horizon cutoff, $L = R_E$, and pressureless matter model. In the upper figure we have  $r_0 = 0.43$ and $c^2 = 0.9$ while the parameter $\epsilon$ varies. In the lower figure we have the variation of the three parameters, $r_0$, $\epsilon$ and $c^2$.  In both panels curves overlap, regardless of the values ​​of the parameters used.}
\label{figholcdm2}
\end{figure}

With the above result, where $\epsilon = 1, 2, 3$, we can calculate the optical depth in this particular cutoff. They are shown in Figure \ref{figholcdm2}. Here we can see that the probability of finding gravitational lenses is basically the same when we change the model parameters, and it is difficult to distinguish the curves.
\par
\begin{figure}[tb]
\includegraphics[width=\columnwidth]{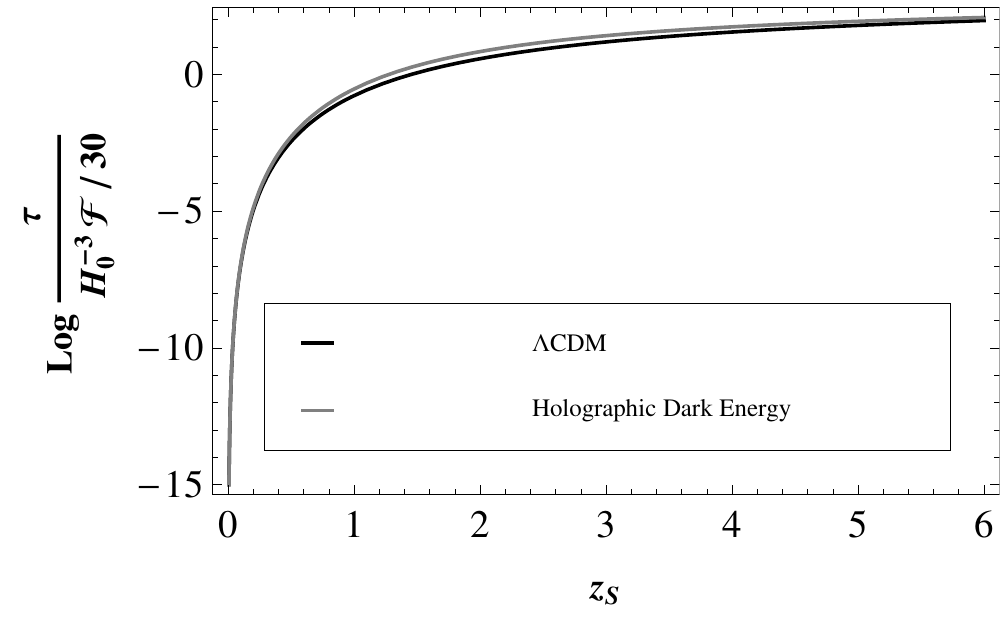}
\caption{Optical depth comparison as a function of the source redshift $z_S$ between the holographic dark energy model, with future event horizon cutoff (solid grey line), $L = R_E$ and $\epsilon = 3$, and the $\Lambda$CDM model (solid black line).}
\label{figholcdm3}
\end{figure}

The situation where the holographic dark energy model is similar to $\Lambda$CDM model is shown in Figure \ref{figholcdm3} and also in this case the superposition of the probability curves is very large but now this situation is indicative of the theoretical similarity of the holographic dark energy model with $L = R_E$ and $\epsilon = 3$ and the $\Lambda$CDM model.

%%%%%%%%%%%%%%%%%%%%%%%%%%%%%%%%%%%%%%%%%%%%%%%%%%
\section{Conclusions}
%%%%%%%%%%%%%%%%%%%%%%%%%%%%%%%%%%%%%%%%%%%%%%%%%%

We studied here the qualitative behavior of the optical depth $\tau$ in terms of some cosmological models.
\par
\begin{figure*}[tb]
%\centering
%\includegraphics[width=\twocolumn]{lentes20}
\includegraphics[width=\textwidth,height=12cm]{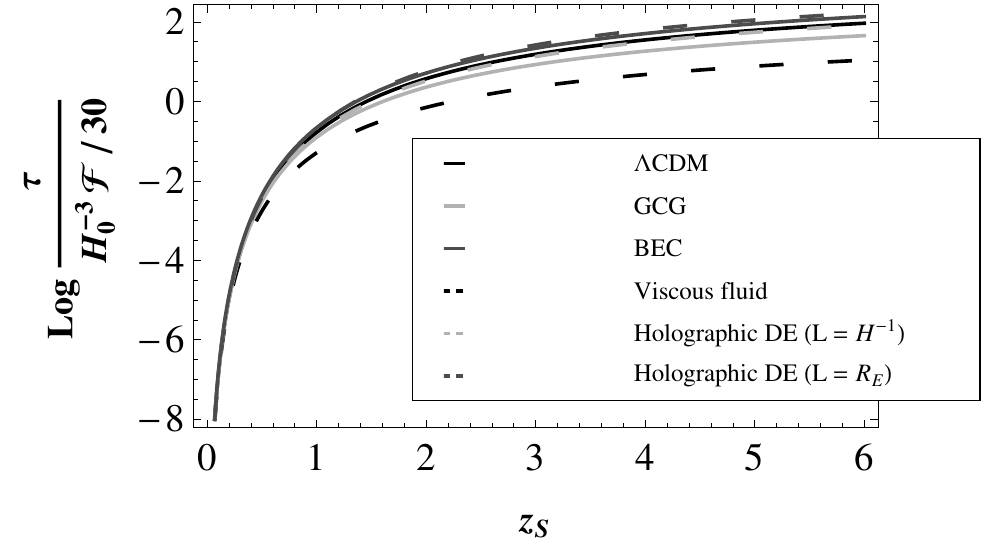}
\caption{Comparing the optical depth as a function of the source redshift $z_S$ in six different models: $\Lambda$CDM ($\Omega_m=0.3$, $\Omega_\Lambda=0.7$), generalized Chaplygin gas ($\Omega_b=0.05$, $\Omega_{Ch}=0.95$, $A_{Ch}=0.7$, $\alpha=0.3$), Bose-Einstein condensate dark matter ($\Omega_b=0.045$, $\Omega_{BEC}=0.255$, $\Omega_{\Lambda}=0.7$, $\omega_{\textrm{BEC}}=10^{3}$), viscous fluid ($\Omega_{m}=0.3$, $\Omega_{vis}=0.7$, $A_{vis}=0.6$, $\beta=0$, $\nu=-3/2$), holographic dark energy with Hubble cutoff ($\Omega_m=0.3$, $\Omega_{Hol}=0.7$, $q_{0}=-0.5$, $n=1/2$) and future event cutoff ($\Omega_m=0.3$, $\Omega_{Hol}=0.7$, $\epsilon=2$, $c^2=0.9$). Here we see more clearly the difficulty in distinguishing between various models with respect to depth optics. We can visualize a difference for large values ​​of $z_S$ in cases of GCG model (solid gray line) and the viscous model (dashed black line). The other four cosmological models overlap. The figure also suggests that the similarities between the cosmological models are very strong for small values ​​of $z_S$.}
\label{figcomp}
\end{figure*}

We can see that, in general, the gravitational lens effects is dependent on the values of the cosmological parameters. This dependence is well illustrated in Figure \ref{figcomp}, although there a small superposition among the models $\Lambda$CDM, BEC, viscous fluid and holographic dark energy ($ H = L ^ {-1} $) for particular choices of parameters and it is impossible to distinguish them with the simple analysis carried out here.
\par
As we are dealing with cosmological models quite different, it is impossible to make a joint analysis of their behavior regarding the relationship between their free parameters and how their variations can infuencethe optical depth in the study of gravitational lensing. Next, we discuss each cosmological model separately and when there is a similarity between the models, we will make a joint analysis.
\par
In the CDM model (Figure \ref{fig_cdm1}) the statistical probability obtained shows that the optical depth $\tau$ decreases with the increasing density parameter and is greater in a Universe with low density. The situation is different in the $\Lambda$CDM model, Figure \ref{fig_lcdm1}. The statistical probability of finding gravitational lenses increases when we increase the amount of the cosmological constant in this model. It is higher when there is no dark matter as part of the material content of the Universe. The BEC model is indifferent to the equation of state when the density parameter $\Omega_{BEC}$ is almost the whole of the material content of the Universe (top figure in Figure \ref{fig_bec1}). This result is confirmed in the middle figure that shows the overlap of the curves in this model on a reduced scale. When the cosmological constant is introduced (figure below in Figure \ref{fig_bec1}), we see that the probability $\tau$ decreases for lower values of the equation of state parameter $\omega_{BEC}$. Furthermore, we can see in Figure \ref{fig_bec2} that the presence of the cosmological constant with the Bose-Einstein condensate dark matter increase the values of the statistical probability of finding gravitational lenses.
\par
For the Chaplygin gas model the statistical probability is very sensitive to changes in model parameters: Large values of $\Omega_{Ch}$ and $A_{Ch}$ along with a small value of $\Omega_{m}$ results in a high value of the optical depth (top figure in Figure \ref{Cgm1}). Another combination of parameters produces a different optical depth curve showing how this model is dependent on the parameters used. In the case of the generalized Chaplygin gas model this dependency is reduced and the differences between the various curves are very small (the lower figure of Fig \ref{Cgm1}).
\par
In general, the viscous fluid model is also very dependent on the values of the parameters used (Figure \ref{figvisc}, Figure \ref{figvisc1} and lower figure in Figure \ref{figvisc2}). However, in the upper figure in Figure \ref{figvisc2} we see a superposition of curves when changing parameters of density, with a small predominance of the values ($ \Omega_m = 0.045, \Omega = 0.7) $ over other .
\par
The holographic dark energy model follows the same dependence-model, both for the case of the Hubble-scale cutoff as in the case of future event horizon cutoff, that was analyzed in the previously cosmological models (see Figures \ref{fighol} - \ref{figholcdm3}). When the density parameters are changed in cutting Hubble scale (lower figure in Figure \ref{fighollcdm}) we observed that the higher probability is found when the matter content of the universe is composed of holographic dark energy $\Omega_ {hol}$ with pressureless matter.
\par
There are cosmological models where the increasing of the amount of dark matter (baryonic and non-baryonic) results in a lower optical depth, namely the CDM model, the GCG model and the viscous cosmological model (see Figure \ref{fig_cdm1}, Figure \ref{Cgm1} and Figure \ref{figvisc2}, upper figure). The situation occurs when the Universe is younger because the optical depth is almost the same for $z\approx 0$. This may seem strange at first given that in a Universe with only dark matter the opposite should happen. There is here, from our point of view, the combination of two situations. In the CDM model the production of gravitational lenses decreases in a situation where there are more baryonic and non-baryonic matter in a less volume that corresponds to a younger universe. On the other hand, in models in which the dark energy is a component of the material content of the Universe, the interaction between dark matter and dark energy changes the cosmological scenario producing the results found by us (see \citep{silvia}, pg 80 for a similar situation). Anyway this kind of behavior should be tested at some point with the observational data to confirm its validity.
\par
The phenomenological approach developed here should of course be complemented by lens models more realistic than provided by SIS model like the Navarro, Frenk and White density profile \citep{NFW00, NFW01}, lens models which are embedded in an external shear field, which is created by matter in the neighbourhood or models obtained by adding two more parameters: the ellipticity and the position angle describing the orientation of the lens. With this, we can make a qualitative analysis and calculate the expected number of gravitational lenses produced by cosmological models studied here and then compare these results with the observed gravitational lenses \citep{tur01, tur02, fuku}. A more detailed comparison between the theoretical predictions for gravitational lensing and observational data should be made in order to restrict more strongly these cosmological models. We will explore these possibilities in future works.

\acknowledgments
We would like to express our gratitude to  C.P. Constantidinis, J.L. Gonzales and C.G.F. Silva for the helpful suggestions and for a careful reading of the manuscript. We are grateful to the referee for his useful comments. This work has received partial financial supporting from CNPq (Brazil), CAPES (Brazil) and FAPES (Brazil).

\end{document}